\documentclass[aps,prresearch,onecolumn,superscriptaddress]{revtex4-2}

\usepackage{graphicx}
\usepackage{amsmath,amssymb}
\usepackage{hyperref}
\usepackage{tikz}
\usetikzlibrary{arrows.meta}
\usepackage{float}
\usepackage{subcaption}
\usepackage{booktabs}
\usepackage{listings}
\usepackage{xcolor}

\begin{document}

\title{Topological Obstructions for Quantum Adiabatic Algorithms: Evidence from MaxCut Instances}

\author{Prathamesh S. Joshi}
\affiliation{Department of Computer Science, Yeshiva University}

\begin{abstract}
Quantum adiabatic algorithms are commonly analyzed through local spectral properties of an interpolating Hamiltonian, most notably the minimum energy gap. While this perspective captures an important constraint on adiabatic runtimes, it does not fully describe the global structure of spectral evolution in optimization problems with degenerate solution manifolds. In this work, we show that degeneracy alone imposes unavoidable global constraints on spectral flow, even in instances where adiabatic algorithms succeed with high probability.

Focusing on digitized quantum adiabatic evolutions, we analyze the eigenphases of the cumulative unitary operator generated along the interpolation path. By explicitly tracking eigenphase trajectories, we demonstrate that multiple spectral bands are forced to interact, braid, and permute before coalescing into a degenerate manifold at the end of the evolution. This global reordering manifests as persistent spectral congestion and nontrivial band permutations that cannot be removed by increasing evolution time or refining the digitization.

Using MaxCut instances with controlled degeneracy as a concrete setting, we extract quantitative diagnostics of spectral congestion and explicitly compute the induced band permutations. Our results show that successful adiabatic optimization can coexist with complex and constrained spectral flow, revealing a form of topological obstruction rooted in the global connectivity of eigenstates rather than in local gap closures. These findings highlight intrinsic limitations of gap-based analyses and motivate spectral-flow–based diagnostics for understanding adiabatic algorithms in degenerate optimization landscapes.
\end{abstract}

\maketitle

\section{Introduction}

Quantum adiabatic algorithms (QAA) provide a conceptually simple framework for solving optimization problems~\cite{farhi2000quantum,albash2018adiabatic} by encoding solutions into the extremal eigenspaces of a problem Hamiltonian and preparing these states through continuous or discretized interpolation from a simple initial Hamiltonian. Since their introduction, the performance of adiabatic algorithms has been closely associated with spectral properties of the interpolating Hamiltonian, most notably the minimum energy gap encountered along the evolution. Small gaps are widely regarded as indicators of increased algorithmic difficulty and longer required runtimes.~\cite{jansen2007bounds,avron1987adiabatic}.

While gap-based reasoning captures an important aspect of adiabatic dynamics, it does not fully describe the structure of spectral evolution in optimization problems with degenerate solution manifolds. In many combinatorial problems, including MaxCut~\cite{goemans1995improved}, multiple distinct configurations achieve the optimal objective value. As a result, the target Hamiltonian exhibits a degenerate extremal eigenspace, and multiple spectral branches must be continuously connected into this manifold as the interpolation parameter is varied. This requirement imposes global constraints on spectral evolution that are not reflected in local gap estimates.

In practice, adiabatic algorithms often succeed on such degenerate instances, producing high-probability samples of optimal solutions even when the spectrum exhibits extended regions of crowding or near-degeneracy. This apparent coexistence of algorithmic success and spectral complexity raises a fundamental question: what aspects of spectral structure truly constrain adiabatic evolution, beyond the existence of isolated small gaps?

In this work, we address this question by shifting attention from local spectral features to the global organization of spectral flow. Rather than focusing on instantaneous Hamiltonian spectra, we analyze the eigenphases of the cumulative unitary operator ~\cite{zhou2020quantum} generated along a digitized adiabatic interpolation. This unitary perspective allows us to track how spectral bands evolve, interact, and ultimately reorganize as the interpolation parameter varies.

Our central observation is that degeneracy at the target Hamiltonian enforces unavoidable global reordering of spectral bands. Even in idealized, noiseless settings and even when adiabatic evolution succeeds with near-unit probability, eigenphase trajectories exhibit persistent congestion and nontrivial permutations before coalescing into the degenerate solution manifold. These features cannot be removed by simply increasing the total evolution time or refining the digitization, indicating a structural obstruction rooted in the global connectivity of eigenstates along the interpolation path. \cite{HastingsFreedman2013}

To make these ideas concrete, we study digitized quantum adiabatic evolutions for MaxCut instances with controlled degeneracy. By explicitly tracking eigenphase bands, quantifying spectral congestion, and extracting the induced band permutations, we provide direct numerical evidence of global spectral constraints that are invisible to standard gap-based diagnostics. While we do not identify quantized topological invariants, we show that the observed obstructions are topological in a pragmatic sense: they reflect global features of spectral connectivity that persist under smooth deformations of the adiabatic path.

The remainder of this paper is organized as follows. In Sec.~II, we introduce the unitary spectral-flow framework and clarify our notion of topological obstruction. In Sec.~III, we describe the digitized-QAA harness and spectral diagnostics used in our simulations. Sec.~IV presents numerical results across multiple MaxCut instances and discretization depths. In Sec.~V, we discuss the implications of these findings for adiabatic diagnostics and optimization algorithms, and we conclude in Sec.~VI.

\section{Spectral Flow in Quantum Adiabatic Evolution}
\label{sec:spectral_flow}

\subsection{From Hamiltonian Interpolation to Unitary Spectral Flow}

Quantum adiabatic algorithms are most often formulated in terms of a parameter-dependent Hamiltonian
\begin{equation}
H(s) = (1-s)\,H_M + s\,H_C , \qquad s \in [0,1],
\label{eq:H_interp}
\end{equation}
whose extremal eigenspaces interpolate between an easily preparable initial state and the solution manifold of an optimization problem. Analyses of adiabatic performance traditionally focus on the instantaneous spectrum of \(H(s)\), with particular emphasis on the minimum energy gap encountered along the path.

In circuit-based implementations and numerical simulations, adiabatic evolution is naturally realized in a digitized form. The continuous interpolation is approximated by a sequence of short evolutions generated alternately by \(H_M\) and \(H_C\), producing a cumulative unitary operator ~\cite{trotter1959product, Yi2021}
\begin{equation}
U(0 \to s_k) = \prod_{\ell=1}^{k} e^{-i \beta_\ell H_M} e^{-i \gamma_\ell H_C},
\label{eq:U_cumulative}
\end{equation}
where the parameters \(\beta_\ell, \gamma_\ell\) follow a discretized schedule approximating the continuous path.

This formulation shifts attention from the instantaneous Hamiltonian spectrum to the spectrum of the cumulative unitary operator \(U(0 \to s)\). Its eigenvalues,
\begin{equation}
\mathrm{spec}\big(U(0 \to s)\big) = \{ e^{i\theta_j(s)} \},
\end{equation}
lie on the unit circle and evolve continuously with the interpolation parameter. The trajectories \(\theta_j(s)\) define the \emph{unitary spectral flow}. Importantly, qualitative features of adiabatic evolution, such as avoided crossings, level repulsion, and state reorganization, remain encoded in this unitary description, but now appear as global properties of eigenphase motion rather than local energy gaps.

\subsection{Degeneracy and Global Constraints on Spectral Evolution}

A central complication in optimization problems is the presence of degeneracy in the target Hamiltonian \(H_C\). In problems such as MaxCut, multiple distinct configurations often achieve the optimal objective value, leading to a degenerate extremal eigenspace at \(s=1\).

From the perspective of spectral flow, degeneracy imposes a strong global constraint. Multiple eigenstates originating from distinct regions of the initial spectrum must be continuously connected into the same degenerate manifold as \(s \to 1\). These connections cannot be arranged independently: spectral branches must interact, exchange character, and reorganize along the path. As a result, the evolution generically exhibits extended regions of spectral congestion and repeated avoided crossings.

Crucially, these features are not tied to algorithmic failure. As we will show numerically, digitized adiabatic evolutions can converge to the correct solution manifold with near-unit probability while still displaying highly nontrivial spectral flow. This observation highlights a distinction between \emph{successful state preparation} and \emph{spectral simplicity}.

\subsection{Topological Obstruction as Global Spectral Reordering}

In this work, we use the term \emph{topological obstruction} to refer to global constraints on spectral evolution that arise from the connectivity of eigenstates under continuous deformation.

Specifically, an obstruction is present when there exists no globally smooth labeling of eigenstates along the interpolation path. In the unitary setting, this manifests as unavoidable reordering of eigenphase bands: when bands are tracked continuously from \(s \approx 0\) to \(s = 1\), their final ordering differs from their initial ordering, defining a nontrivial permutation. Such permutations reflect global features of the spectral flow that cannot be eliminated by rescaling the total evolution time or refining the discretization.

This notion of obstruction is topological in the sense that it depends on global connectivity rather than on local spectral features. ~\cite{avron1983homotopy} It captures the fact that, in degenerate optimization landscapes, spectral branches are forced to braid and permute before merging into the solution manifold.

\subsection{Schematic Illustration of Spectral Flow}

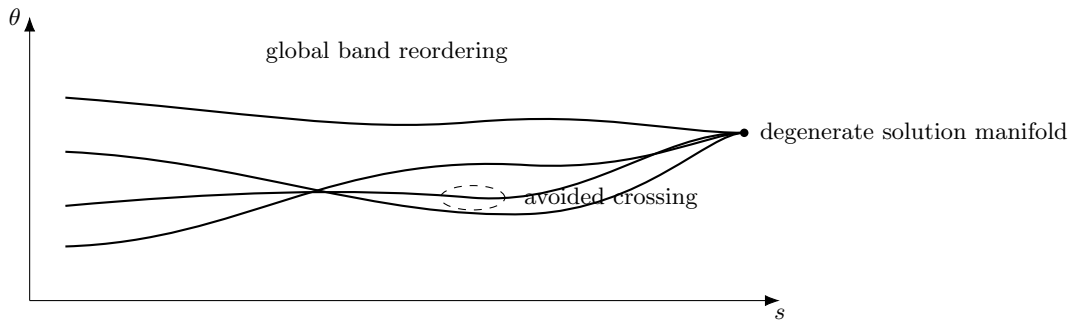
\begin{figure}[t]
\centering
\begin{tikzpicture}[x=9.5cm,y=3.6cm]
  \draw[-{Latex[length=2mm]}] (0,0) -- (1.05,0) node[below] {$s$};
  \draw[-{Latex[length=2mm]}] (0,0) -- (0,1.05) node[left] {$\theta$};

  \draw[thick] (0.05,0.20)
    .. controls (0.30,0.22) and (0.42,0.55) .. (0.70,0.50)
    .. controls (0.86,0.48) and (0.95,0.62) .. (1.00,0.62);

  \draw[thick] (0.05,0.35)
    .. controls (0.26,0.40) and (0.45,0.42) .. (0.62,0.38)
    .. controls (0.78,0.35) and (0.88,0.62) .. (1.00,0.62);

  \draw[thick] (0.05,0.55)
    .. controls (0.30,0.52) and (0.48,0.30) .. (0.70,0.32)
    .. controls (0.86,0.34) and (0.95,0.62) .. (1.00,0.62);

  \draw[thick] (0.05,0.75)
    .. controls (0.30,0.70) and (0.45,0.62) .. (0.62,0.66)
    .. controls (0.80,0.70) and (0.90,0.62) .. (1.00,0.62);

  \fill (1.00,0.62) circle (1.6pt);
  \node[anchor=west] at (1.01,0.62) {degenerate solution manifold};

  \draw[dashed] (0.62,0.38) circle (0.045);
  \node[anchor=west] at (0.68,0.38) {\small avoided crossing};

  \node at (0.50,0.92) {\small global band reordering};
\end{tikzpicture}
\caption{Schematic (illustration) of unitary spectral flow in a degenerate optimization problem. As the interpolation parameter \(s\) varies, eigenphase bands interact and exchange ordering before coalescing into a degenerate solution manifold at \(s=1\). The resulting global reordering motivates the spectral diagnostics introduced in this work.}
\label{fig:schematic_flow}
\end{figure}

Figure~\ref{fig:schematic_flow} provides a schematic illustration of the spectral-flow perspective adopted in this work. As the interpolation parameter \(s\) varies, multiple eigenphase bands interact through avoided crossings and exchange ordering. At the end of the evolution, several bands must coalesce into a degenerate manifold corresponding to the optimal solutions.

This illustration is not intended to represent a specific instance or numerical result. Rather, it highlights the qualitative features - spectral congestion, band interaction, and global reordering that motivate the diagnostics developed in the remainder of the paper. In later sections, we make these ideas precise by explicitly tracking eigenphase bands and extracting the induced permutations in concrete MaxCut instances.

\section{MaxCut as a Testbed for Spectral Obstruction}
\label{sec:maxcut}

\subsection{Problem Definition and Degenerate Solution Structure}

The MaxCut problem is a canonical combinatorial optimization task defined on an undirected graph \(G=(V,E)\). Given a binary assignment \(z_i \in \{0,1\}\) to each vertex \(i \in V\), the cost function counts the number of edges whose endpoints are assigned different values,
\begin{equation}
C(z) = \sum_{(i,j)\in E} \frac{1}{2}\left(1 - z_i z_j\right),
\end{equation}
up to an unimportant affine transformation. The objective is to find assignments maximizing \(C(z)\).

A key feature of MaxCut, and the primary reason for its relevance here, is the frequent presence of multiple optimal solutions. For many graphs, the maximum cut value is achieved by several distinct bitstrings related by symmetries or frustration in the graph structure. Consequently, the problem Hamiltonian encoding MaxCut typically possesses a degenerate extremal eigenspace.

This degeneracy plays a central role in the spectral behavior of adiabatic algorithms. From the perspective of spectral flow, multiple eigenstates must converge into a single degenerate manifold as the interpolation parameter approaches its final value. As discussed in Section~\ref{sec:spectral_flow}, this requirement constrains how spectral branches can evolve.

\subsection{Hamiltonian Encoding and Digitized Adiabatic Evolution}

We encode MaxCut into a cost Hamiltonian of the form
\begin{equation}
H_C = \sum_{(i,j)\in E} \frac{1}{2}\left( \mathbb{I} - Z_i Z_j \right),
\label{eq:HC}
\end{equation}
whose lowest-energy eigenspace corresponds to optimal cuts. The mixing Hamiltonian is chosen as
\begin{equation}
H_M = \sum_{i \in V} X_i,
\label{eq:HM}
\end{equation}
which has a nondegenerate ground state given by the uniform superposition over all computational basis states.

Rather than implementing continuous-time evolution under an interpolating Hamiltonian, we employ a digitized adiabatic protocol. The evolution is approximated by a sequence of alternating unitaries generated by \(H_M\) and \(H_C\),
\begin{equation}
U(0 \to s_k) = \prod_{\ell=1}^{k} e^{-i \beta_\ell H_M} e^{-i \gamma_\ell H_C},
\end{equation}
with linearly varying parameters \(\beta_\ell\) and \(\gamma_\ell\) chosen to approximate the interpolation in Eq.~\eqref{eq:H_interp}. This construction is closely related to Trotterized adiabatic evolution and differs from variational QAOA in that the parameters are fixed by an explicit schedule rather than optimized \cite{Binkowski2023}.

The digitized setting is particularly convenient for our purposes, as it provides direct access to the cumulative unitary operator \(U(0 \to s)\) at intermediate stages of the evolution. This allows us to study the spectral flow of eigenphases without invoking instantaneous Hamiltonian spectra.

\subsection{Successful Optimization Despite Spectral Congestion}

For the graph instances studied in this work, the digitized adiabatic protocol successfully prepares the MaxCut solution manifold with high probability. In numerical simulations, the final measurement distributions are sharply peaked on the optimal bitstrings, and the total probability mass on the optimal set approaches unity as the discretization is refined.

At the level of output statistics alone, the algorithm appears to perform as expected. However, this apparent simplicity contrasts sharply with the behavior of the spectral flow observed during the evolution. As we demonstrate explicitly in Section~\ref{sec:results}, the eigenphases of the cumulative unitary operator exhibit pronounced congestion, repeated near-collisions, and nontrivial reordering throughout the interpolation.

This coexistence of algorithmic success and spectral complexity is central to the present work. It underscores the fact that reaching the correct solution manifold does not imply a simple or weakly interacting spectral structure. On the contrary, the spectral flow reflects global constraints imposed by degeneracy and connectivity, even when the final outcome is favorable.

\subsection{Why MaxCut Is Sufficient for Revealing Obstruction}

The MaxCut instances considered here are small enough to be simulated exactly, yet they already display all the structural ingredients relevant to spectral obstruction: degeneracy, frustration, and competing spectral branches. Our goal is not to claim hardness or to analyze asymptotic scaling, but to expose concrete mechanisms by which global spectral reordering arises in adiabatic optimization.

By focusing on MaxCut, we isolate these mechanisms in a setting where they can be visualized, tracked, and quantified directly. In the following section, we introduce diagnostics based on eigenphase tracking and band permutation, and we show how these reveal robust, nontrivial spectral reorganization across different discretizations and graph instances.

\section{Numerical Protocol and Spectral Diagnostics}
\label{sec:methods}

\subsection{Digitized Adiabatic Protocol}

All numerical experiments in this work are performed using a digitized adiabatic evolution, implemented as a sequence of alternating unitaries generated by the mixing and cost Hamiltonians defined in Eqs.~\eqref{eq:HM} and~\eqref{eq:HC}. The cumulative unitary at interpolation step \(k\) is given by
\begin{equation}
U_k = \prod_{\ell=1}^{k} e^{-i \beta_\ell H_M} e^{-i \gamma_\ell H_C},
\label{eq:Uk}
\end{equation}
where the parameters follow a linear schedule,
\begin{equation}
\beta_\ell = (1 - s_\ell)\,\Delta t, \qquad
\gamma_\ell = s_\ell\,\Delta t, \qquad
s_\ell = \frac{\ell}{K}.
\end{equation}
Here \(K\) denotes the number of Trotter steps and \(\Delta t = T/K\) the step size for total evolution time \(T\).

The initial state is prepared as the uniform superposition over computational basis states, corresponding to the ground state of \(H_M\). Final measurement statistics are obtained by sampling the computational basis after application of the full unitary \(U_K\).

\subsection{Extraction of Unitary Spectral Flow}

At selected values of the interpolation parameter \(s_k\), we compute the eigenvalues of the cumulative unitary \(U_k\),
\begin{equation}
U_k \, |\phi_j^{(k)}\rangle = e^{i\theta_j(s_k)} |\phi_j^{(k)}\rangle,
\end{equation}
and record the eigenphases \(\theta_j(s_k) \in (-\pi,\pi]\). The resulting trajectories \(\theta_j(s)\) define the unitary spectral flow discussed in Section~\ref{sec:spectral_flow}.

To quantify spectral congestion, we compute the minimum adjacent phase spacing
\begin{equation}
\Delta\theta_{\min}(s_k) = \min_j \left| \theta_{j+1}(s_k) - \theta_j(s_k) \right|,
\end{equation}
after sorting phases at fixed \(s_k\). Persistent suppression of \(\Delta\theta_{\min}\) across the interpolation indicates unavoidable spectral congestion.

\subsection{Band Tracking and Permutation Diagnostics}

To probe global reorganization of eigenstates, we track a subset of eigenphase bands continuously from \(s \approx 0\) to \(s = 1\). Band tracking is performed by maximizing overlap between eigenvectors at consecutive steps,
\begin{equation}
\mathcal{O}_{ij}^{(k)} = \left| \langle \phi_i^{(k)} | \phi_j^{(k+1)} \rangle \right|,
\end{equation}
and assigning bands according to maximal overlap. This procedure yields continuous eigenphase trajectories even in the presence of avoided crossings.

Comparing the ordering of tracked bands at the beginning and end of the interpolation defines a permutation \(\pi\) of eigenphase indices. Nontrivial cycle structure in \(\pi\) indicates global band reordering and the absence of a globally smooth eigenstate labeling along the adiabatic path.

\subsection{Illustrative Implementation}

The essential structure of the digitized evolution and spectral extraction is illustrated by the following representative code fragment:

\begin{lstlisting}[language=Python, caption={Overlap-based band tracking procedure}]
for k in range(K):
    beta = (1 - k/K) * dt
    gamma = (k/K) * dt

    qc = QuantumCircuit(n)
    for q in range(n):
        qc.rx(mixer_scale * beta, q)
    for (i, j) in edges:
        qc.rzz(-2 * gamma, i, j)

    U = Operator(qc).data
    phases, vecs = np.linalg.eig(U)
    record_phases(phases, vecs, k)
\end{lstlisting}

This snippet highlights the essential ingredients: cumulative unitary construction, eigenphase extraction, and storage of spectral data for subsequent analysis. Full implementation details, including histogram generation, band tracking, and permutation analysis, follow this structure and are used consistently across all graph instances studied.

\subsection{Graph Instances and Parameters}

We apply this protocol to multiple MaxCut instances with varying graph sizes and structures. For each graph, we perform simulations at several discretization depths \(K\) while keeping the total evolution time \(T\) fixed. This allows us to distinguish features that persist under refinement of the digitization from those attributable to finite-step artifacts.

In the next section, we present the resulting spectral flow data, permutation statistics, and output distributions, and demonstrate how nontrivial spectral obstruction persists even when the adiabatic algorithm succeeds with near-unit probability.

\section{Numerical Results}
\label{sec:results}

\subsection{Successful Optimization with Degenerate Outputs}

We begin by verifying that the digitized adiabatic protocol successfully prepares the MaxCut solution manifold for all graph instances studied. For each graph, we compute the exact optimal cut value and its degeneracy via brute-force enumeration. The final measurement distribution obtained after application of the full unitary \(U_K\) is then compared against this optimal set.

\begin{figure}[H]
  \centering
  \begin{subfigure}[t]{0.32\textwidth}
    \centering
    \includegraphics[width=\linewidth]{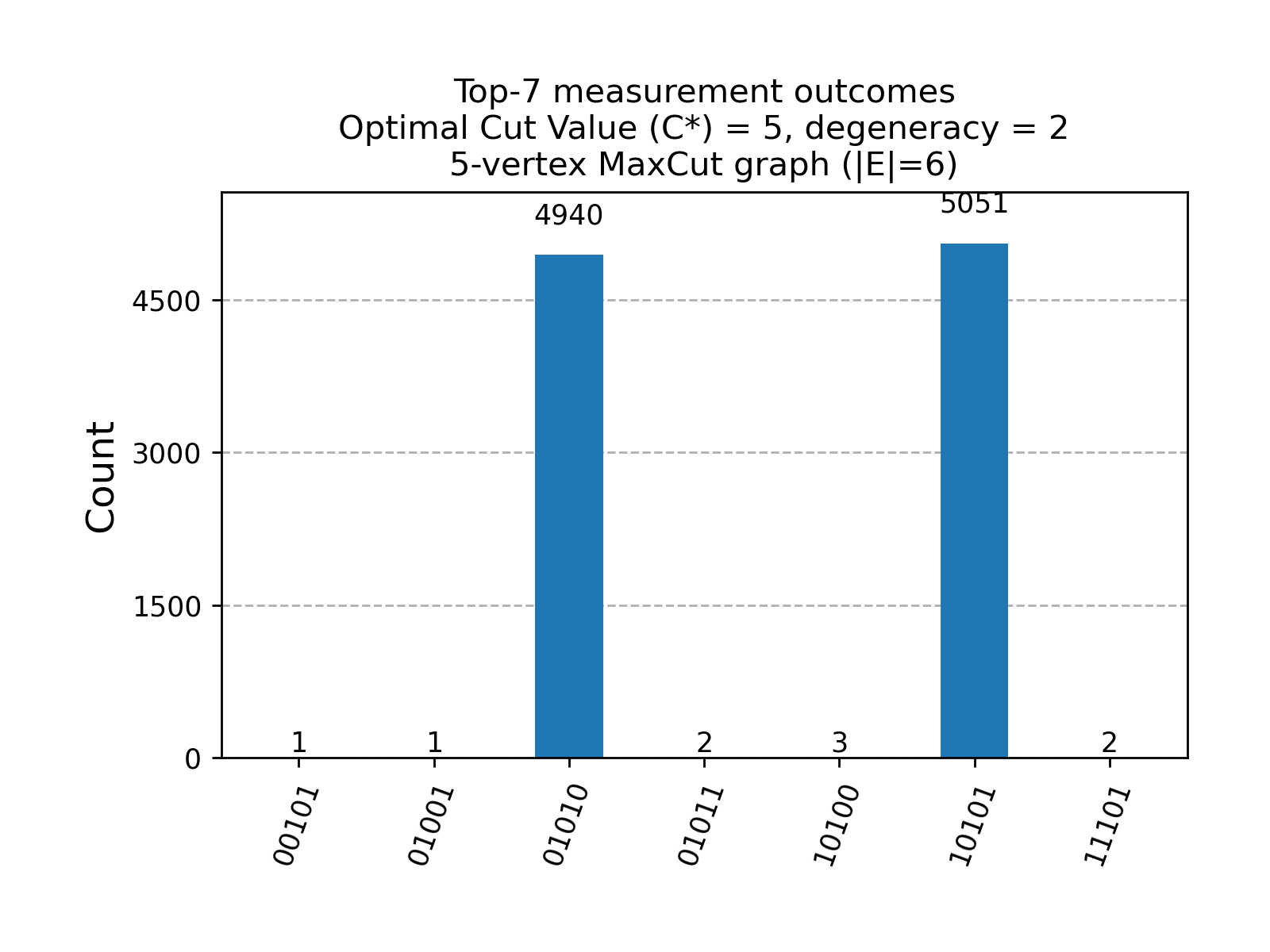}
    \caption{$n=5,\ K=160$}
  \end{subfigure}\hfill
  \begin{subfigure}[t]{0.32\textwidth}
    \centering
    \includegraphics[width=\linewidth]{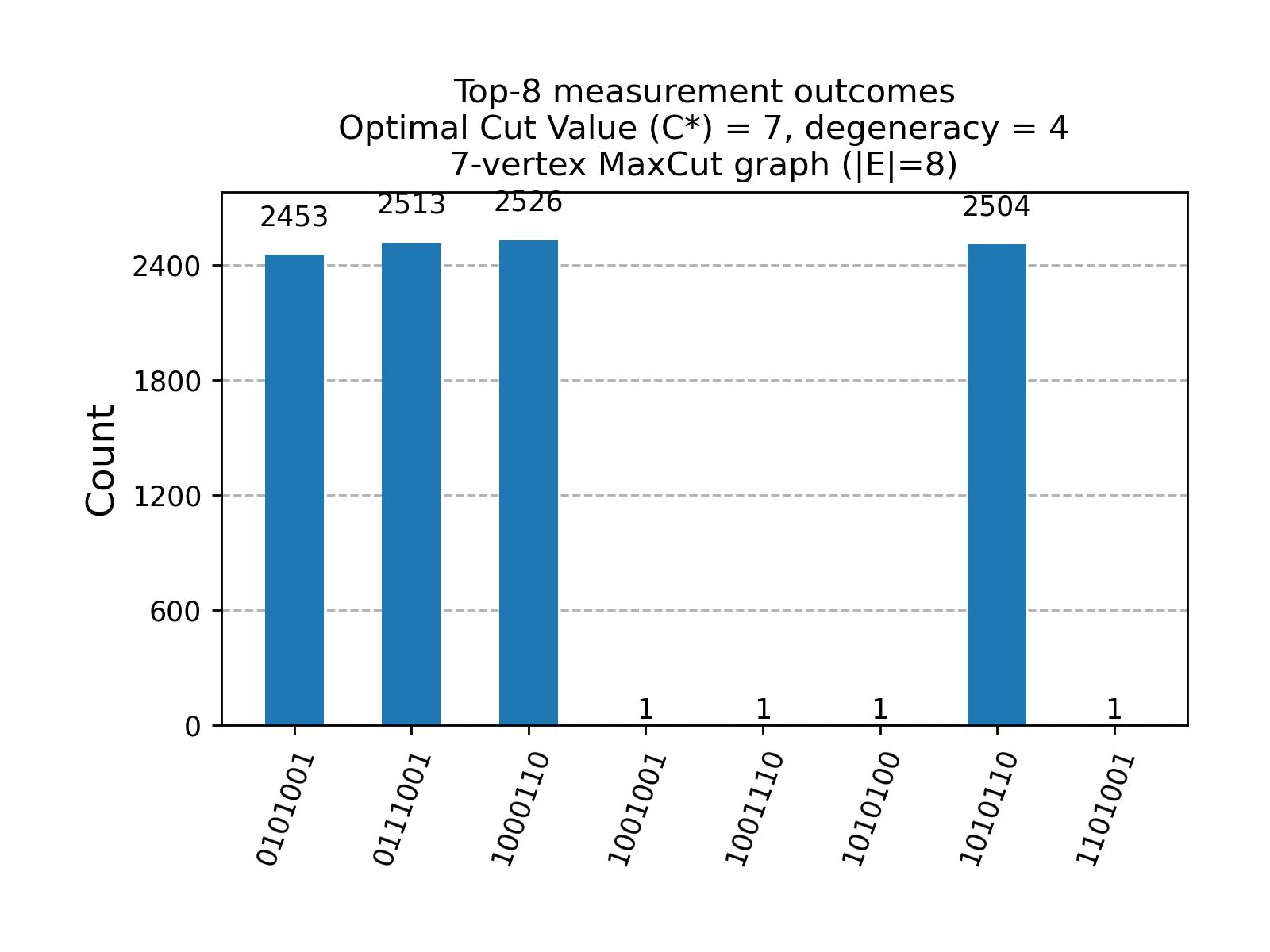}
    \caption{$n=7,\ K=160$}
  \end{subfigure}\hfill
  \begin{subfigure}[t]{0.32\textwidth}
    \centering
    \includegraphics[width=\linewidth]{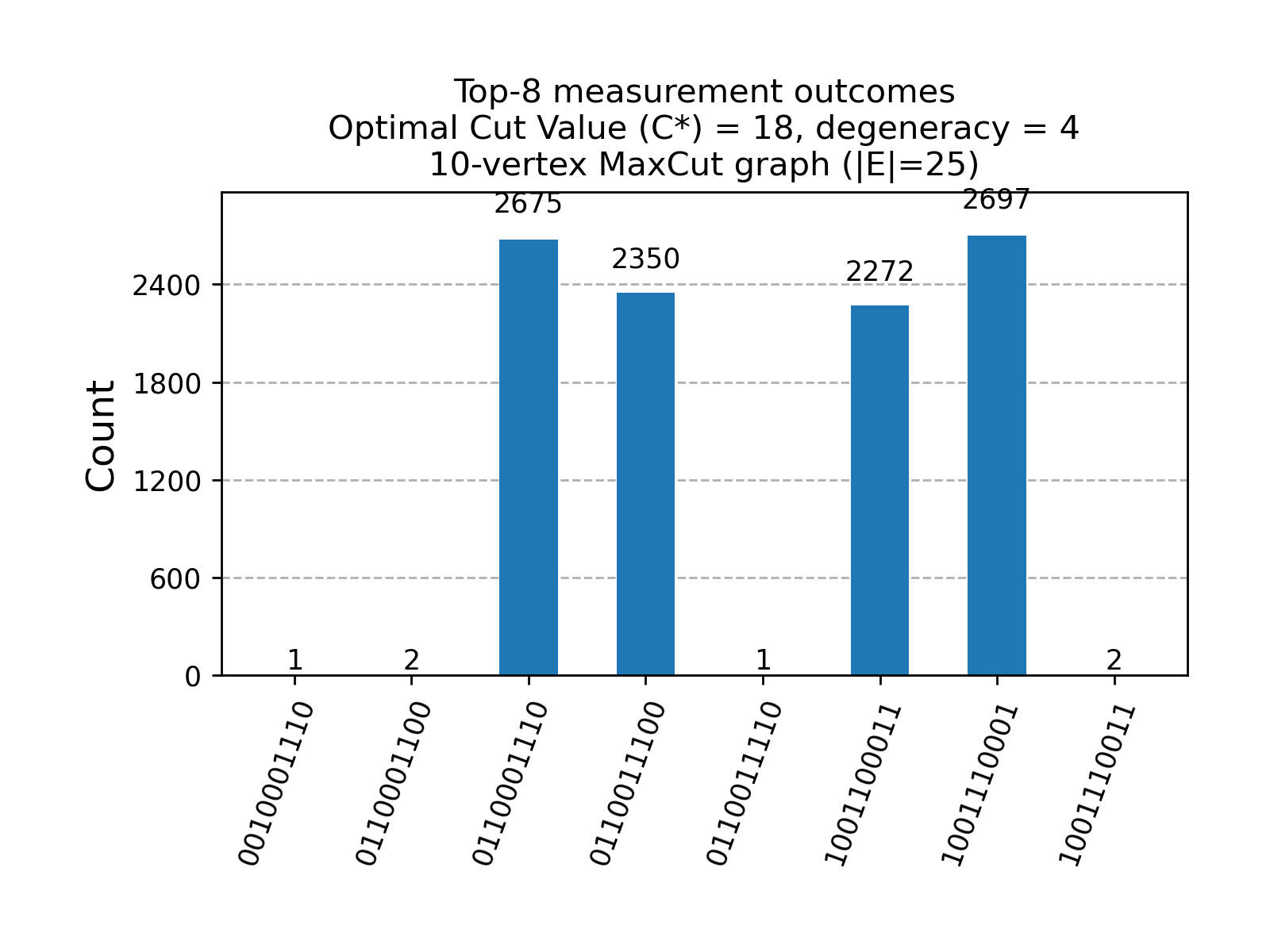}
    \caption{$n=10,\ K=240$}
  \end{subfigure}

  \caption{Representative final-time measurement histograms (top outcomes), showing concentration on the degenerate optimal MaxCut solutions across different graph sizes.}
  \label{fig:histograms}
\end{figure}

Across all instances and discretization depths considered, the probability mass on the optimal solution set approaches unity as the number of Trotter steps \(K\) is increased, while keeping the total evolution time \(T\) fixed. Representative histograms for selected graphs are shown in Fig.~\ref{fig:histograms}. In each case, the dominant measurement outcomes correspond precisely to the degenerate optimal bitstrings.

These results confirm that the digitized adiabatic evolution succeeds in its algorithmic objective: the correct solution manifold is reached with high probability, and no evidence of diabatic failure is observed at the level of output statistics.

\subsection{Persistent Spectral Congestion Along the Adiabatic Path}

\begin{figure}[H]
  \centering

  \begin{subfigure}[t]{0.49\textwidth}
    \centering
    \includegraphics[width=\linewidth]{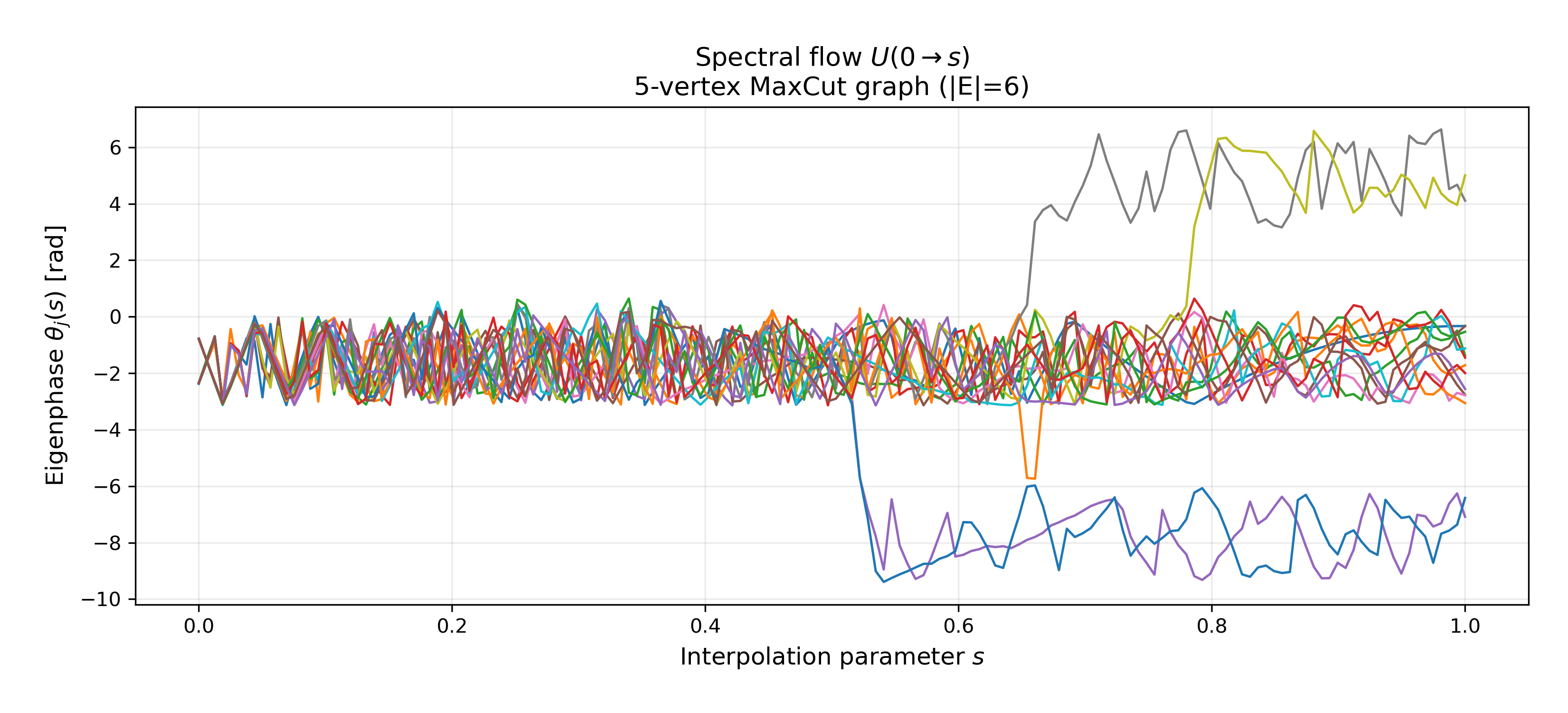}
    \caption{$n=5,\ K=160$}
  \end{subfigure}
  \hfill
  \begin{subfigure}[t]{0.49\textwidth}
    \centering
    \includegraphics[width=\linewidth]{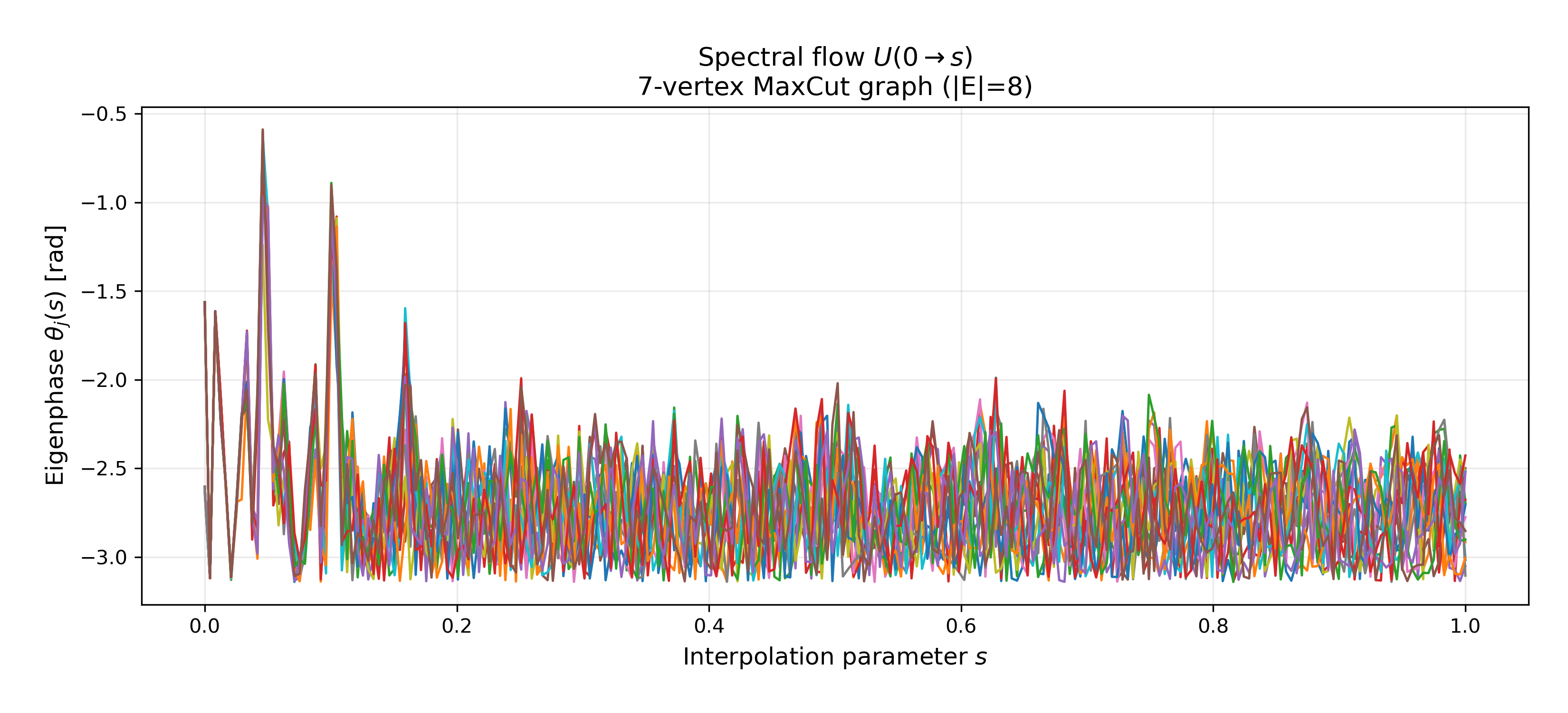}
    \caption{$n=7,\ K=240$}
  \end{subfigure}

  \vspace{0.1em}

  \begin{subfigure}[t]{0.49\textwidth}
    \centering
    \includegraphics[width=\linewidth]{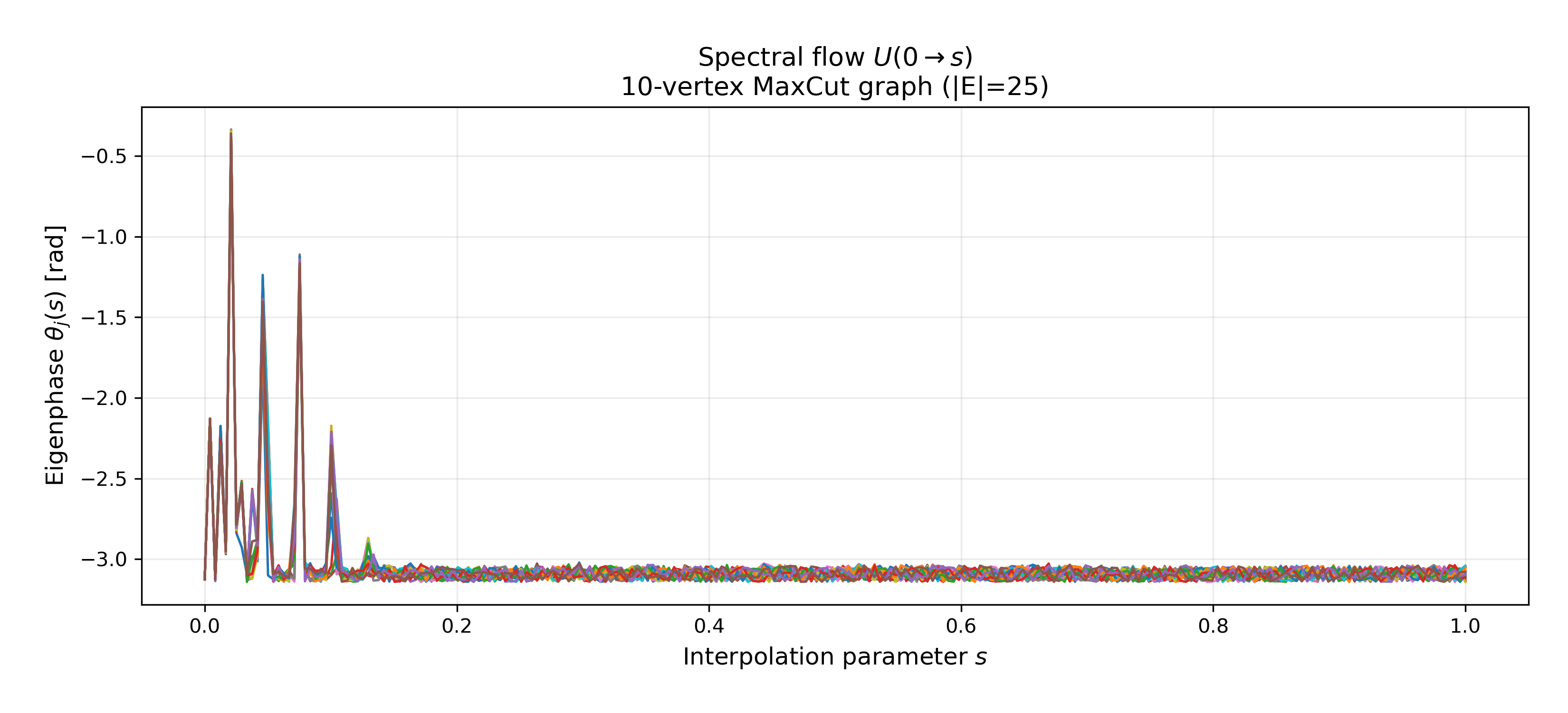}
    \caption{$n=10,\ K=240$}
  \end{subfigure}
  \hfill
  \begin{subfigure}[t]{0.49\textwidth}
    \centering
    \includegraphics[width=\linewidth]{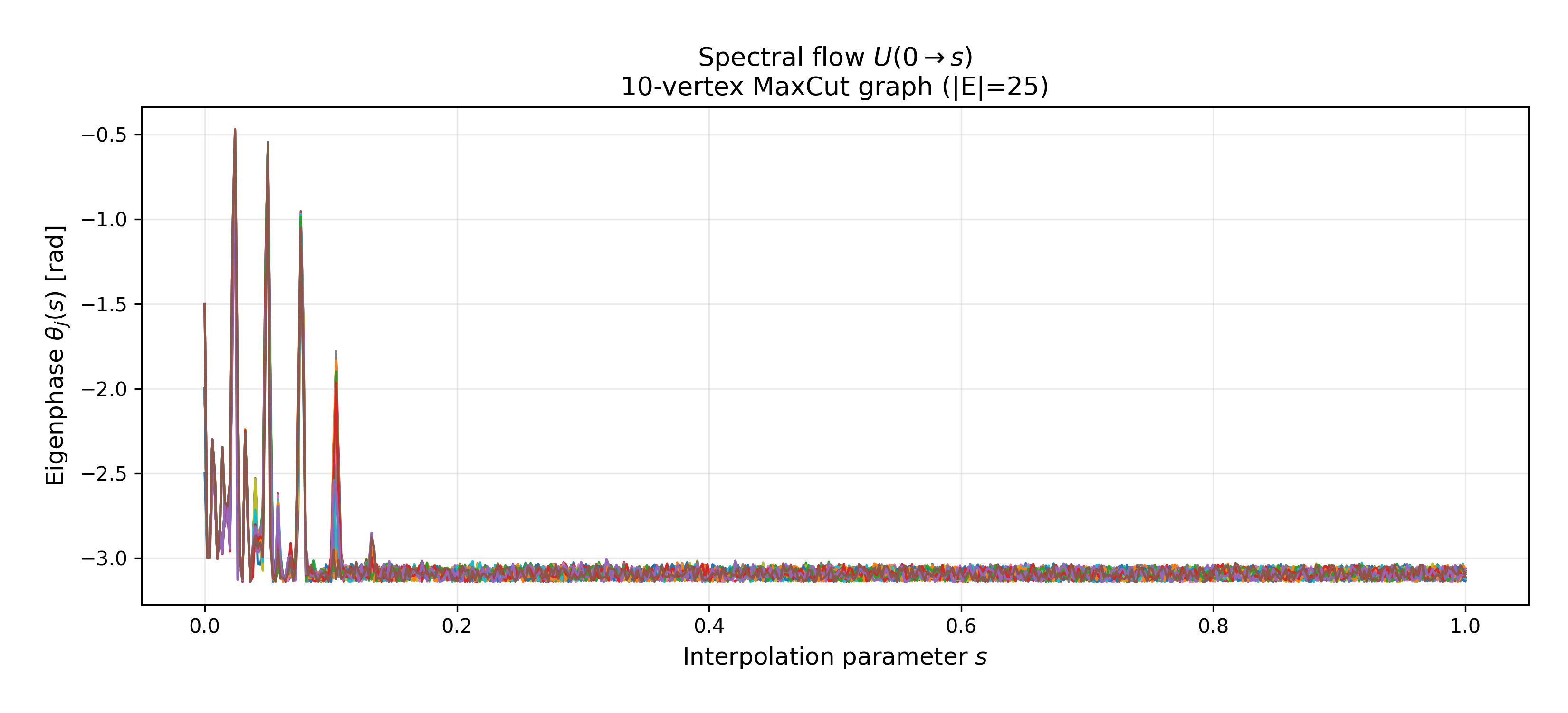}
    \caption{$n=10,\ K=500$}
  \end{subfigure}

  \caption{Unitary spectral flow $U(0\to s)$ (eigenphases $\theta_j(s)$) shown across problem sizes and digitization depth.}
  \label{fig:spectral_flow}
\end{figure}

\begin{figure}[H]
  \centering

  \begin{subfigure}[t]{0.49\textwidth}
    \centering
    \includegraphics[width=\linewidth]{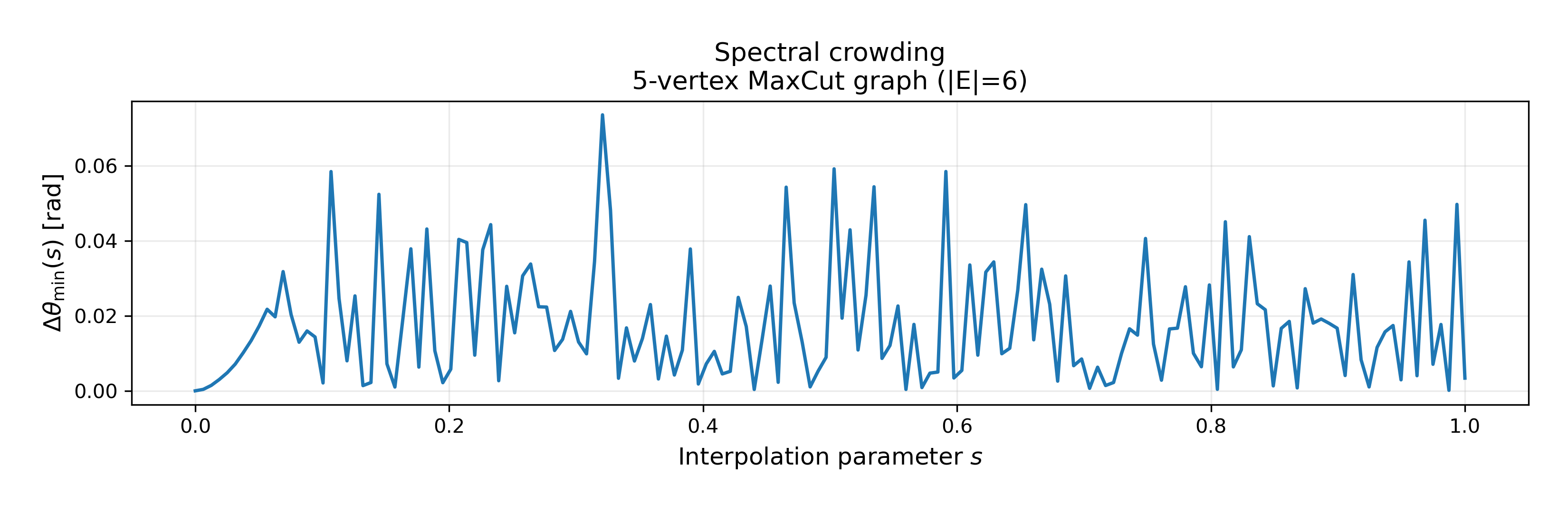}
    \caption{$n=5,\ K=160$}
  \end{subfigure}
  \hfill
  \begin{subfigure}[t]{0.49\textwidth}
    \centering
    \includegraphics[width=\linewidth]{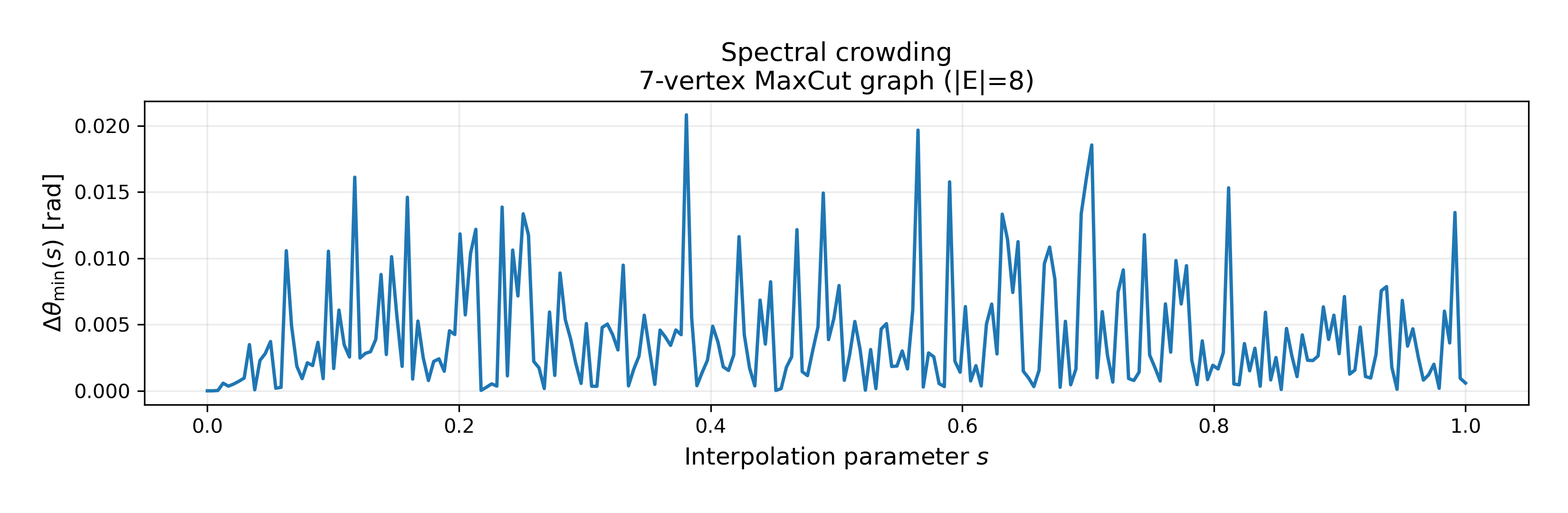}
    \caption{$n=7,\ K=240$}
  \end{subfigure}

  \vspace{0.1em}

  \begin{subfigure}[t]{0.49\textwidth}
    \centering
    \includegraphics[width=\linewidth]{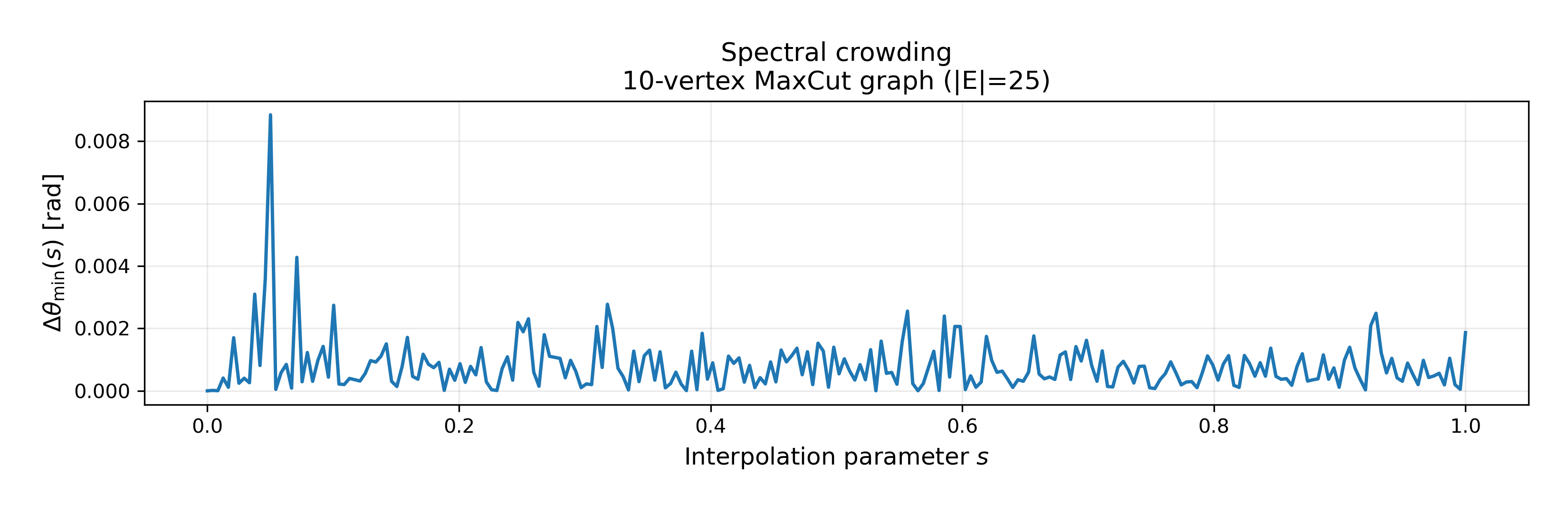}
    \caption{$n=10,\ K=240$}
  \end{subfigure}
  \hfill
  \begin{subfigure}[t]{0.49\textwidth}
    \centering
    \includegraphics[width=\linewidth]{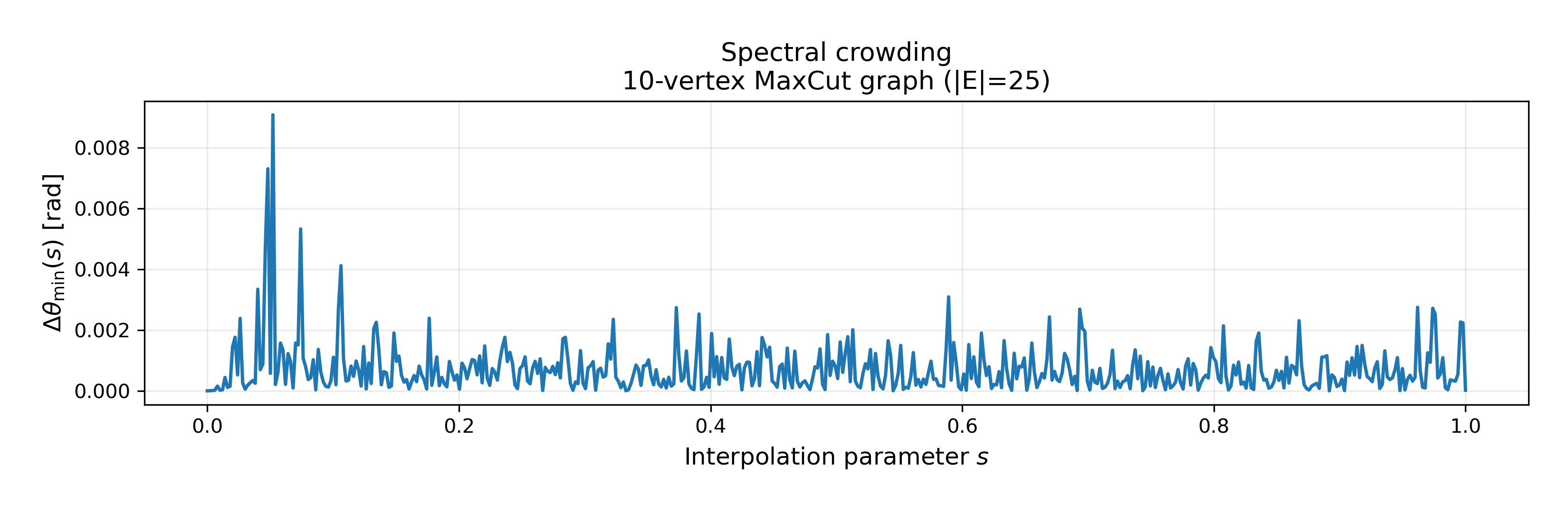}
    \caption{$n=10,\ K=500$}
  \end{subfigure}

  \caption{spectral congestion diagnostic $\Delta\theta_{\min}(s)$ (minimum eigenphase spacing) for the same instances shown in Fig.~\ref{fig:spectral_flow}.}
  \label{fig:crowding}
\end{figure}

Despite the apparent simplicity of the final measurement outcomes, the
eigenphase evolution of the cumulative unitary $U(0\to s)$ exhibits
substantial structure throughout the interpolation.
Figure~\ref{fig:spectral_flow} displays representative eigenphase trajectories
for several MaxCut instances and discretization depths, demonstrating that
the adiabatic path is generically accompanied by extended regions of strong
spectral interaction.

For the smallest instance ($n=5$), the eigenphase bands organize into a small
number of visibly separated groups, with intermittent interactions but
relatively clear separations between bundles.
As the problem size increases to $n=7$, this structure becomes noticeably
more congested: eigenphase bands repeatedly approach one another over wide
intervals of the interpolation, producing sustained regions of local
spectral congestion.
For the largest instance studied ($n=10$), this effect is amplified.
A large fraction of the eigenphases remain confined within a narrow phase
window over most of the interpolation, forming a densely packed spectral
region that persists even as the digitization depth is increased from
$K=240$ to $K=500$.

To quantify this behavior, we introduce the minimum adjacent eigenphase
spacing
\[
\Delta\theta_{\min}(s) = \min_j \bigl| \theta_{j+1}(s) - \theta_j(s) \bigr|,
\]
computed after sorting the eigenphases at each value of $s$ (with circular
wrap-around).
This quantity provides a direct diagnostic of spectral congestion.
Figure~\ref{fig:crowding} shows $\Delta\theta_{\min}(s)$ for the same instances
appearing in Fig.~\ref{fig:spectral_flow}.

Across all graphs and discretization depths examined, $\Delta\theta_{\min}(s)$
is repeatedly suppressed over extended intervals of the interpolation,
indicating frequent near-collisions between eigenphase bands.
Importantly, refining the digitization does not remove this effect.
While increasing $K$ sharpens fine-scale features and slightly shifts the
locations of individual minima, the overall scale and persistence of spectral
crowding remain essentially unchanged.
In particular, for the $n=10$ instance, $\Delta\theta_{\min}(s)$ remains orders
of magnitude smaller than for $n=5$ throughout the evolution, even at the
largest discretization depth considered.

These observations demonstrate that spectral congestion is not a numerical
artifact of coarse Trotterization.
Rather, it reflects a structural property of the adiabatic path induced by
the presence of a degenerate solution manifold.
As multiple eigenphase branches are forced to merge into the same terminal
subspace, repeated close encounters between bands become unavoidable along
the interpolation.
This establishes that high adiabatic success probability can coexist with
persistent local spectral congestion.

\subsection{Global Band Reordering and Permutation Structure}

While the diagnostics in Section~V.B establish the presence of persistent
local spectral congestion, they do not by themselves characterize how
eigenphase bands connect \emph{globally} from the beginning to the end of the
adiabatic interpolation.
To probe this global organization, we explicitly track a fixed subset of
eigenphase bands continuously from $s \approx 0$ to $s = 1$ using the
overlap-based procedure described in Section~\ref{sec:methods}.

\begin{table}[H]
\centering
\begin{tabular}{llrrrrrr}
\toprule
\textbf{Instance} & \textbf{$K$} &
\textbf{$C^\star$} & \textbf{deg.} &
\textbf{$P_{\mathrm{succ}}$} &
\textbf{median $\Delta\theta_{\min}$} &
\textbf{max $\Delta\theta_{\min}$} &
\textbf{\# nontrivial cycles} \\
\midrule

$n=5,\ |E|=6$\\
& 160$^{\ast}$ & 5 & 2 & 0.9991 & 0.01382 & 0.07356 & 2 \\
& 240 & 5 & 2 & 0.9997 & 0.01187 & 0.07435 & 3 \\
& 500 & 5 & 2 & 0.9995 & 0.01261 & 0.07368 & 3 \\
\midrule

$n=7,\ |E|=8$\\
& 160 & 7 & 4 & 0.9996 & 0.002752 & 0.02184 & 2 \\
& 240$^{\ast}$ & 7 & 4 & 0.9998 & 0.002729 & 0.02082 & 3 \\
& 500 & 7 & 4 & 0.9999 & 0.002557 & 0.01826 & 4 \\
\midrule

$n=10,\ |E|=25$\\
& 160 & 18 & 4 & 0.9972 & $5.61\times 10^{-4}$ & 0.007960 & 3 \\
& 240$^{\ast}$ & 18 & 4 & 0.9994 & $5.77\times 10^{-4}$ & 0.008847 & 4 \\
& 500$^{\ast}$ & 18 & 4 & 0.9996 & $4.80\times 10^{-4}$ & 0.009089 & 2 \\
\bottomrule
\end{tabular}

\vspace{0.5em}
\begin{minipage}{0.95\linewidth}
\footnotesize
$^{\ast}$Runs marked with an asterisk are shown explicitly in Fig.~3.
The final column reports the number of nontrivial cycles (length $\geq 2$) in the
end-to-end permutation of eigenphase bands induced by overlap-based tracking
from $s=0$ to $s=1$.
\end{minipage}

\caption{Run-level summary for digitized-QAA MaxCut experiments at fixed $T=50$
(mixer scale $=5$), across three graph instances and discretization depths
$K\in\{160,240,500\}$. We report the exact optimum cut value $C^\star$, its
degeneracy, empirical success probability $P_{\mathrm{succ}}$, and spectral
crowding statistics from the cumulative eigenphase flow via
$\Delta\theta_{\min}(s)$ (median and maximum over $s\in[0,1]$).}
\label{tab:run_summary}
\end{table}

Comparing the ordering of the tracked eigenphase bands at the start and end of
the interpolation defines a permutation of band indices.
If a globally smooth labeling of eigenstates existed along the adiabatic
path, this permutation would be trivial.
A nontrivial permutation, in particular, one containing cycles of length
$\geq 2$---therefore provides a direct signature of global band reordering
that cannot be inferred from local gap or crowding information alone.

Table~\ref{tab:run_summary} summarizes the resulting permutation structure for
all graph instances and discretization depths studied.
Across all cases, the induced permutation contains multiple nontrivial
cycles, indicating that eigenphase bands are globally rearranged between
$s=0$ and $s=1$.
Although the detailed cycle decomposition varies with the instance and the
value of $K$, the presence of nontrivial cycles is universal.

Notably, increasing the number of Trotter steps does not eliminate this
global reordering.
In several cases, finer discretization resolves additional cycle structure
that is obscured at coarser $K$, rather than restoring a trivial permutation.
This demonstrates that the observed band reordering is not a discretization
artifact, but a robust feature of the adiabatic evolution.

Taken together, the results of Sections~V.B and V.C show that local spectral
congestion and global band reordering coexist throughout the adiabatic path.
Although eigenphases evolve continuously as functions of $s$, their
connectivity enforces a nontrivial end-to-end permutation before the bands
coalesce into the degenerate solution manifold at $s=1$.
This end-to-end reordering constitutes a \emph{global spectral obstruction}:
it does not prevent adiabatic success, but it cannot be removed by local
refinements of the interpolation or by increasing the digitization depth \cite{Cheon1998}.

\subsection{Robustness Across Graph Instances and Discretization}

We observe consistent qualitative behavior across graph sizes ranging from five to ten vertices and across multiple graph topologies. Smaller graphs exhibit fewer bands and larger typical phase spacings, while larger graphs display denser spectra and more intricate reordering. Nevertheless, the core phenomena - spectral congestion, repeated near-collisions, and nontrivial band permutation, are present in all cases.

Crucially, these features coexist with near-perfect algorithmic success. The results therefore highlight a separation between optimization performance and spectral simplicity: the adiabatic algorithm can succeed even while the underlying spectral flow remains globally constrained and topologically nontrivial.

\subsection{Summary of Observations}

Taken together, the numerical results establish three central observations:
\begin{enumerate}
    \item Digitized quantum adiabatic evolution successfully prepares degenerate MaxCut solution manifolds with high probability.
    \item The associated unitary spectral flow exhibits persistent congestion that is insensitive to discretization refinement.
    \item Eigenphase bands undergo unavoidable global reordering, captured by nontrivial permutation structure, indicating the absence of a globally smooth eigenstate labeling.
\end{enumerate}

These observations provide concrete evidence that topological obstruction, understood as global spectral reorganization, is an intrinsic feature of quantum adiabatic optimization in degenerate landscapes. In the following section, we discuss the implications of these findings for the interpretation of adiabatic algorithms and their limitations beyond gap-based analyses.

\section{Discussion and Implications}
\label{sec:discussion}

\subsection{Spectral Obstruction Versus Algorithmic Success}

The numerical results presented in Section~\ref{sec:results} demonstrate a clear separation between two notions that are often implicitly conflated in analyses of quantum adiabatic algorithms: algorithmic success and spectral simplicity. In all MaxCut instances studied, the digitized adiabatic evolution successfully prepares the degenerate solution manifold with high probability. At the same time, the associated unitary spectral flow exhibits persistent congestion and nontrivial global reordering of eigenphase bands.

This coexistence highlights an important point. The absence of diabatic failure at the level of output statistics does not imply a simple or weakly interacting spectral structure along the adiabatic path. Instead, even successful evolutions may be forced to navigate highly constrained spectral landscapes. From this perspective, spectral obstruction should be understood as a property of the adiabatic path itself, rather than as a diagnostic of algorithmic failure.

\subsection{Nature of the Obstruction}

It is important to clarify the sense in which the obstruction identified here is ``topological.'' No quantized invariant is computed, nor is a sharp phase transition invoked. Rather, the obstruction arises from global connectivity constraints on the spectrum under continuous deformation. The presence of degeneracy in the target Hamiltonian forces multiple spectral branches to merge at the end of the interpolation, and this requirement induces unavoidable band interactions and reordering along the path.

In the unitary setting, this manifests as the absence of a globally smooth labeling of eigenstates from \(s \approx 0\) to \(s = 1\). The resulting band permutations are robust under refinement of the discretization and rescaling of the total evolution time, indicating that they are not artifacts of Trotterization or numerical noise. In this physically meaningful sense, the obstruction reflects global spectral structure rather than local features.

\subsection{Relation to Gap-Based Analyses}

Traditional analyses of adiabatic algorithms emphasize the minimum spectral gap as the primary determinant of runtime. ~\cite{young2010first} While gap considerations remain essential for understanding adiabatic scaling, the present results suggest that they provide an incomplete picture in degenerate optimization landscapes.

Spectral congestion and band permutation capture information that is invisible to gap-based diagnostics. In particular, they reveal how multiple eigenstates interact globally even when no single isolated gap dominates the evolution. This observation may help explain why certain problem instances exhibit complex adiabatic dynamics despite appearing benign under local spectral measures.

\subsection{Generality Beyond MaxCut}

Although our numerical study focuses on MaxCut, the mechanisms identified here are not specific to this problem. Any optimization task whose solution space is degenerate, either due to symmetry, frustration, or structural redundancy, will impose similar global constraints on spectral evolution. From this standpoint, MaxCut serves as a minimal testbed rather than a special case.

The diagnostics introduced in this work, including eigenphase tracking and permutation analysis, are readily applicable to other adiabatic protocols and problem Hamiltonians. They provide a framework for investigating spectral obstruction in settings where traditional diagnostics are insufficient.

\subsection{Outlook}

The results presented here suggest several directions for future work. One natural extension is to study how spectral obstruction evolves with system size and problem structure, and whether its severity correlates with known notions of computational hardness. Another is to explore the interaction between spectral obstruction and noise, particularly in fault-tolerant or error-corrected implementations where unitary spectral structure may interact nontrivially with error channels.

More broadly, our findings motivate a shift in perspective: rather than asking only whether an adiabatic algorithm succeeds, it may be equally important to ask how the spectrum is forced to reorganize in order for that success to occur. Understanding these global constraints may prove essential for developing a more complete theory of quantum adiabatic optimization.

\section{Conclusion}
\label{sec:conclusion}

In this work, we examined quantum adiabatic optimization from a
spectral-flow perspective, emphasizing the global organization of
eigenphase evolution rather than solely local gap properties.
Using digitized adiabatic protocols applied to MaxCut instances with
degenerate solution manifolds, we showed that high-probability adiabatic
success can coexist with pronounced spectral congestion and unavoidable
global reordering of eigenstates.

By explicitly tracking the eigenphases of the cumulative evolution
operator, we demonstrated that multiple spectral bands are forced to
interact and permute before merging into the degenerate solution space at
the end of the interpolation.
This behavior persists under refinement of the digitization and rescaling
of the total evolution time, indicating that it reflects intrinsic
constraints imposed by degeneracy rather than numerical artifacts.

These results answer the question posed in the Introduction: local spectral
gaps alone do not fully characterize the constraints governing adiabatic
evolution in degenerate optimization problems.
Instead, global connectivity of eigenstates enforces nontrivial end-to-end
band permutations, implying the absence of a globally smooth labeling
along the adiabatic path.

We use the term \emph{topological obstruction} in a pragmatic sense,
referring to such global constraints on spectral connectivity rather than
to quantized topological invariants.
In this sense, the obstruction does not prevent adiabatic success, but it
cannot be removed by local refinements of the interpolation or by
increasing the digitization depth.

More broadly, our findings highlight the value of spectral-flow–based
diagnostics as a complement to traditional gap-based analyses.
They motivate further investigation of global spectral structure in larger
systems and across other classes of optimization problems, where
degeneracy and frustration are expected to play a central role.

\section*{Acknowledgments}

I thank Professor Emil Prodan for suggesting this project and for valuable guidance throughout its development.

\section*{References}
\nocite{*}
\bibliographystyle{apsrev4-2}
\bibliography{references}

\end{document}